\documentclass[%
 preprint, 
 amsmath,amssymb,
 aps, physrev,
]{revtex4-2}
\UseRawInputEncoding
\usepackage{graphicx}
\usepackage{dcolumn}
\usepackage{bm}
\usepackage{xcolor}
\usepackage{soul}
\usepackage{comment} 
\usepackage{hyperref}
\usepackage{hyperref}
\usepackage[mathlines]{lineno}

\begin{document}


\title{\textbf{Selective Parametric Amplification of Degenerate Modes in Electrostatically Transduced Coupled Beam Resonators} 
}%

\author{Vishnu Kumar}
 \email{vishnukumar@iisc.ac.in}
\author{Nishta Arora}
\altaffiliation[Present address: ]{School of Mathematics and Physics, University of Queensland, 4072 Brisbane, Australia.}
\author{Bhargavi B.A.}
\author{Akshay Naik}
\author{Saurabh A. Chandorkar}%
 \email{saurabhc@iisc.ac.in}
\affiliation{%
Centre for Nano Science and Engineering, Indian Institute of Science, 560012 Bengaluru, India.\\
}%


\date{\today}

\begin{abstract}

Parametric excitation in coupled mechanical systems has enabled advances in sensing, computation, and phonon control. The function of distinct phase modes using parametric driving remains insufficiently explored. Here, we investigate the nonlinear and parametric response of degenerate phase modes in a Double Ended Tuning Fork (DETF) resonator. Our measurements reveal pronounced nonlinearity and parametric amplification in the out-phase mode, attributed to the dominant contribution of the coupling beam, while the in-phase mode remains predominantly linear. Uniquely, we demonstrate parametric excitation through the coupling spring, enabling selective amplification and de-amplification controlled via the relative phase between harmonic and parametric drives. A parametric gain of $\sim$13 dB is achieved in the out-phase mode, with phase-dependent modulation of amplification, indicating its suitability for signal processing, logic operations, and memory elements based on degenerate modes. These results establish a new approach to exploiting mode-specific nonlinear dynamics in coupled resonators for emerging applications in sensing and phononic control.

\end{abstract}

\maketitle


\section{\label{sec:level1}Introduction}
Parametric excitation in micro- and nano-electromechanical systems (M/NEMS) provides a powerful tool to control mechanical motion by periodically modulating system parameters, typically stiffness, at twice the resonant frequency \cite{Nishta_parametric}. This approach has been widely used to enhance signal sensitivity and dynamic range in diverse applications, including mass sensing \cite{Turner_mass_sensing}, logic operations \cite{nonlinear_memory}, bifurcation-based amplifiers \cite{Topo_amplifier}, and quantum-limited measurements \cite{Bowen_squeezing_quantum_sensing, quantum_sensing}. In coupled mechanical systems, parametric driving further enables controlled interactions between collective modes, allowing the manipulation of their amplitudes and phases through engineered coupling \cite{mode_coupling_review,Nishta_coupling}. 

Nonlinear effects play a central role in such systems, facilitating signal amplification \cite{high_sensitivity}, noise squeezing \cite{Bowen_para_cooling}, enhanced sensing \cite{Turner_mass_sensing,quantum_sensing,Seshia_para_sensing}, frequency stabilization \cite{Frequency_stability}, and control of dynamical states \cite{Yamaguchi_coherentphonon}. Coupled resonators, in particular, have been shown to exhibit energy transfer \cite{Seshia_coupled_nonlinear}, synchronization \cite{synchronization}, and mode hybridization \cite{mode_hybridization}, offering new ways for coherent phonon control and dynamic mode selection. These effects have spurred the development of high-performance M/NEMS sensors \cite{quantum_sensing,sensing_nano} and computation architectures, including mechanical Ising machines \cite{nonlinear_memory,Chan_Ising}.

Despite these advances, the distinct roles of phase modes, namely in-phase (symmetric) and out-phase (antisymmetric), in governing the parametric dynamics of coupled systems are still not well understood. In particular, the effect of the coupling element on phase mode-specific parametric amplification has received little attention. Prior studies have shown that inter-beam coupling in multimode or membrane systems can modify nonlinearity and energy localization \cite{Seshia_electrostat_modecoupling,Halg_sin_parametric}, however, parametric excitation mediated through the coupling beam remains unexplored.

In this work, we investigate the nonlinear and parametric dynamics of degenerate phase modes in a Double-Ended Tuning Fork (DETF) resonator. Using harmonic and parametric excitation schemes, we directly probe the in-phase and out-of-phase vibrational modes. Remarkably, we observe pronounced nonlinearity and selective parametric amplification in the out-phase mode arising from the dominant contribution of the coupling beam, while the in-phase mode remains largely linear. The selective amplification arises from the coupling beam’s contribution to the
system stiffness, which dominates the nonlinear dynamics. A maximum parametric gain of approximately 13 dB is achieved in the out-phase mode, with clear phase-dependent modulation of amplification. Our findings establish a new approach for mode-selective control of nonlinear dynamics in coupled mechanical systems. This mechanism provides a pathway toward mechanically reconfigurable platforms for signal processing, logic operations, and memory elements, as well as for applications in phononic and quantum transduction \cite{Bagheri_memory,Xiao_phononcavity,Bowen_para_cooling,Villanueva_parametric_oscillator,Samanta2023}.

\section{Design and Measurement Schematic}

\begin{figure}[!h]
    \centering
    \includegraphics[width =\linewidth]{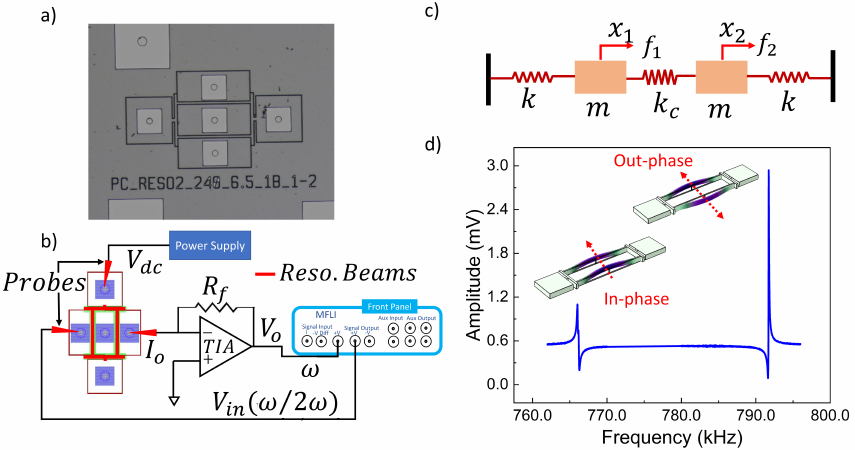}
    \caption{Device measurement technique: (a) Scanning Electron Microscope (SEM) image of the coupled beam resonators, known as DETF resonator, (b) the measurement setup for the nonlinear and parametric measurements using a lock-in amplifier, (c) the spring mass system of DETF resonator (d) the frequency response of the degenerate modes of the DETF, in-phase and out-phase modes.}
    \label{device_setup}
\end{figure}

The DETF resonator is fabricated using the SOI substrate (refer to supplementary section I for detailed fabrication). The SEM image of the fabricated device is shown in \autoref{device_setup}{a}. The first fundamental degenerate modes (in-phase and out-phase) are measured harmonically and parametrically using a Zurich lock-in amplifier (LIA), as shown in \autoref{device_setup}{b}. \autoref{device_setup}{c} shows the spring mass system configuration, highlighting the coupling spring connected to the resonating beams. The degenerate modes arise from the two resonating beams connected through coupling beams. The input excitation is applied on one of the outer electrodes and measured output signal at the other, which is connected to the trans-impedance amplifier with sufficient gain to enhance the signal above the noise level of the electronics and change the current output signal to voltage signal, before sending to the Zurich lock-in amplifier (LIA).  
The frequency response for the degenerate modes is shown in \autoref{device_setup}{d} along with its mode shapes simulated using COMSOL. 

\section{Results and Discussion}

\subsection{Harmonic Response}

\begin{figure}[!h]
    \centering
    \includegraphics[width =\linewidth]{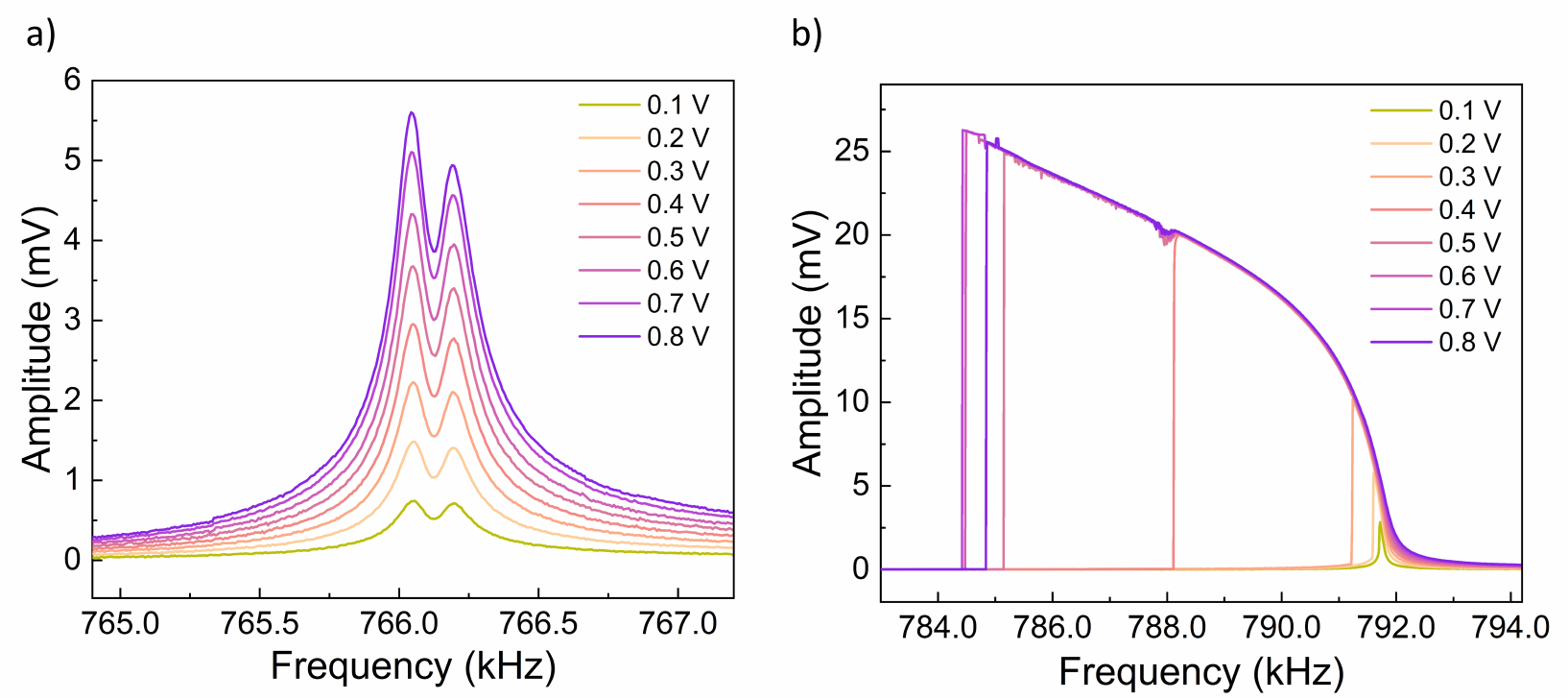}
    \caption{Device harmonic measurement: Frequency response of (a) in-phase mode and (b)out-phase mode  with increasing actuation drive. The out-phase mode shows softening nonlinearity.}
    \label{drive_direct}
\end{figure}

As a coupled system, the DETF consists of two resonating beams linked by coupling beams. At the fundamental frequency, the beams exhibit two degenerate phase modes, an in-phase (IP) mode and an out-phase (OP) mode, with resonant frequencies of 766.1 kHz and 791.75 kHz, respectively. By applying a sinusoidal signal at resonance frequency through electrostatic excitation, the system responds nonlinearly as the excitation increases, and the electrostatic response is measured to certain forces.  
Despite, expecting similar nonlinear response from both the modes, we observed that the out-phase mode is much more nonlinear than the in-phase mode, as shown in \autoref{drive_direct} {a} and \autoref{drive_direct} {b}. 
The in-phase mode also exhibits amplitude splitting due to higher-order mode coupling \cite{Nishta_splitting}. Similarly, the notch in the out-phase response shows the higher-order mode coupling effect on the nonlinearity, is called a surge point \cite{Farbod_nonlinear} and further studies of underlying behavior will be addressed in future work.  
The underlying nonlinear dynamics that attribute to out-phase mode is linked to the strength of the coupling beam. To understand this behavior better, refer analytical modeling in supplementary section II. The equation of motion of two phase modes are:

\begin{equation}\label{eqn 4_20}
    m\Ddot{u}+c\dot{u}+(k_1-f_b)u=f_a cos(\omega t)
\end{equation}
\begin{equation}\label{eqn 4_21}
    m\Ddot{v}+c\dot{v}+(k_1+2k_{c1}-f_b)v-2k_{c3}v^3=f_a cos(\omega t)
\end{equation}

The dynamical behavior of distinct phase modes show the dominant coupling effect that also subjected to the nonlinearity. The softening nonlinear coefficients are in order of $10^{18} N/m^3$ and the increment with excitation shows the linear behavior, as discussed in supplementary section III.

\subsection{Parametric Response}

Parametric excitation induces oscillations in a resonant system by periodically altering one of its parameters, especially the stiffness of the resonator at twice the resonant frequency\cite{Welte_parametric, Dobrindt_parametric}. The two-phase modes are parametrically excited at $2\omega$ and detected at $\omega$ using a lock-in amplifier. The pump signal (drive) is applied from the critical voltage, where the harmonic response begins to exhibit nonlinearity. As excitation increases, the system enters self-oscillation and starts to show a response. This behavior occurs only when the system is in a nonlinear state. The critical pump voltage where the self-oscillation starts is 1.1 V. This critical voltage is required to overcome the damping of the system.

\begin{figure}[!h]
    \centering
    \includegraphics[width =\linewidth]{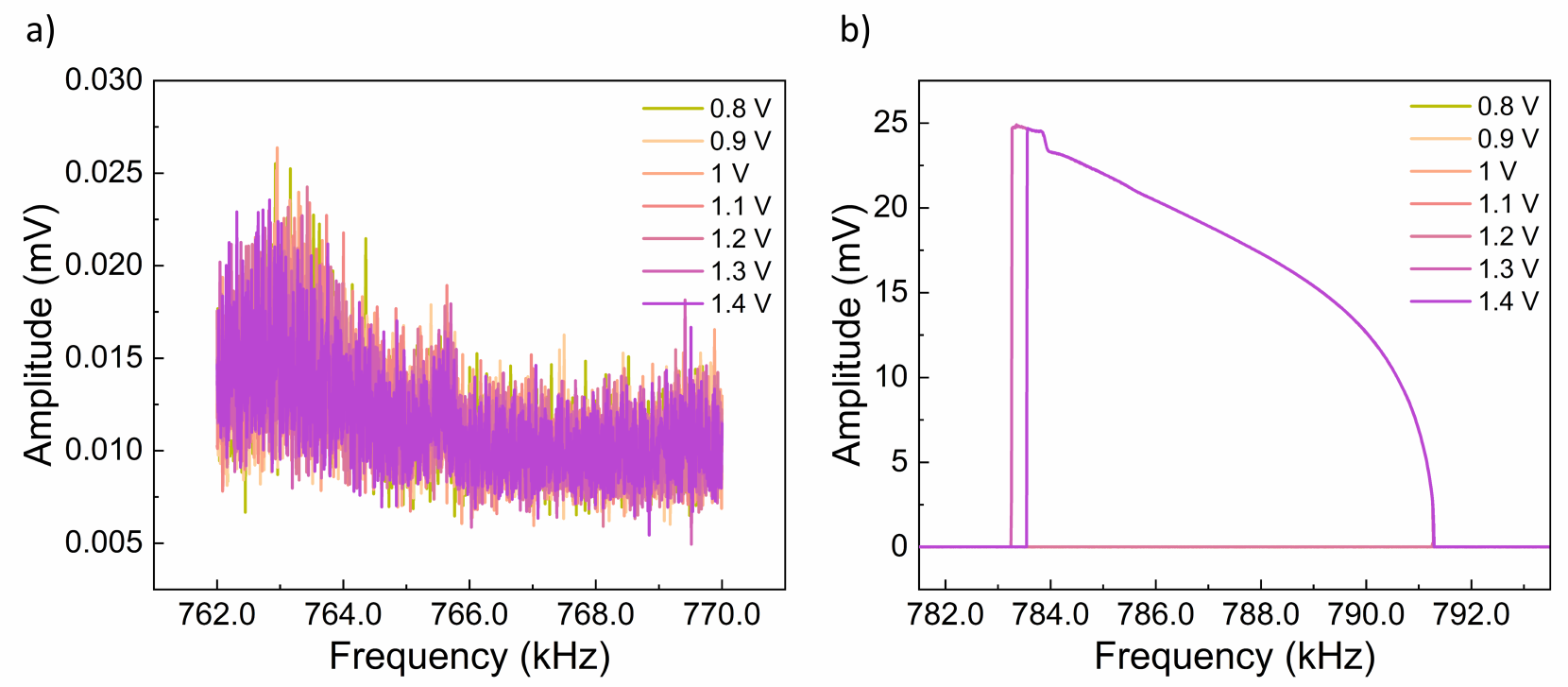}
    \caption{Device parametric measurement: (a) in-phase shows no response, (b) out-phase mode shows nonlinear response from 1.1 V onwards, indicating the coupling beam effect}
    \label{drive_para}
\end{figure}

The response for the in-phase mode shown in \autoref{drive_para} {a} doesn't have any impact of parametric excitation, whereas the out-phase mode response, shown in \autoref{drive_para} {b}, demonstrates parametric action at an excitation of 0.8 V to 1.4 V. The nonlinearity in the reverse sweep comes from the coupling beam contribution, as discussed in the analytical modeling. The equation of motion of out-phase mode showing coupling beam strength effective on the resonating beam, excited parametrically is:

\begin{equation}\label{eqn 4_21}
    m\Ddot{v}+c\dot{v}+(k_1-f_b)v+2k_{c1}(1+\beta sin(2\omega t))v-2k_{c3}v^3=0
\end{equation}

Here, $\beta=\gamma \dfrac{2\epsilon A V_{dc}V_p}{g^2}$, $V_p$ is the parametric excitation voltage. $\gamma$ is a proportionality constant that accounts for the change in coupling spring constant due to application of excitation voltage $V_p$ on the outer electrode that excite beam 1. $k_{c1}$ can be calculated through the in-phase and out-phase mode linear harmonic response as $k_{c1}=0.5*m(\omega_2^2-\omega_1^2)$. The amplitude of the response is governed by the nonlinear term $k_{c3}$.

In the self-oscillation regime for in-phase and out-phase modes, the critical voltage for the out-phase mode is 1.1 V, while the critical voltage for the in-phase mode is much higher ($V_{c,in-phase} \gg V_{c,out-phase}$), as it remains in the linear regime at the same voltage as the out-phase mode.

\subsection{Parametric with Harmonic Response}

\begin{figure}[!h]
    \centering
    \includegraphics[width =\linewidth]{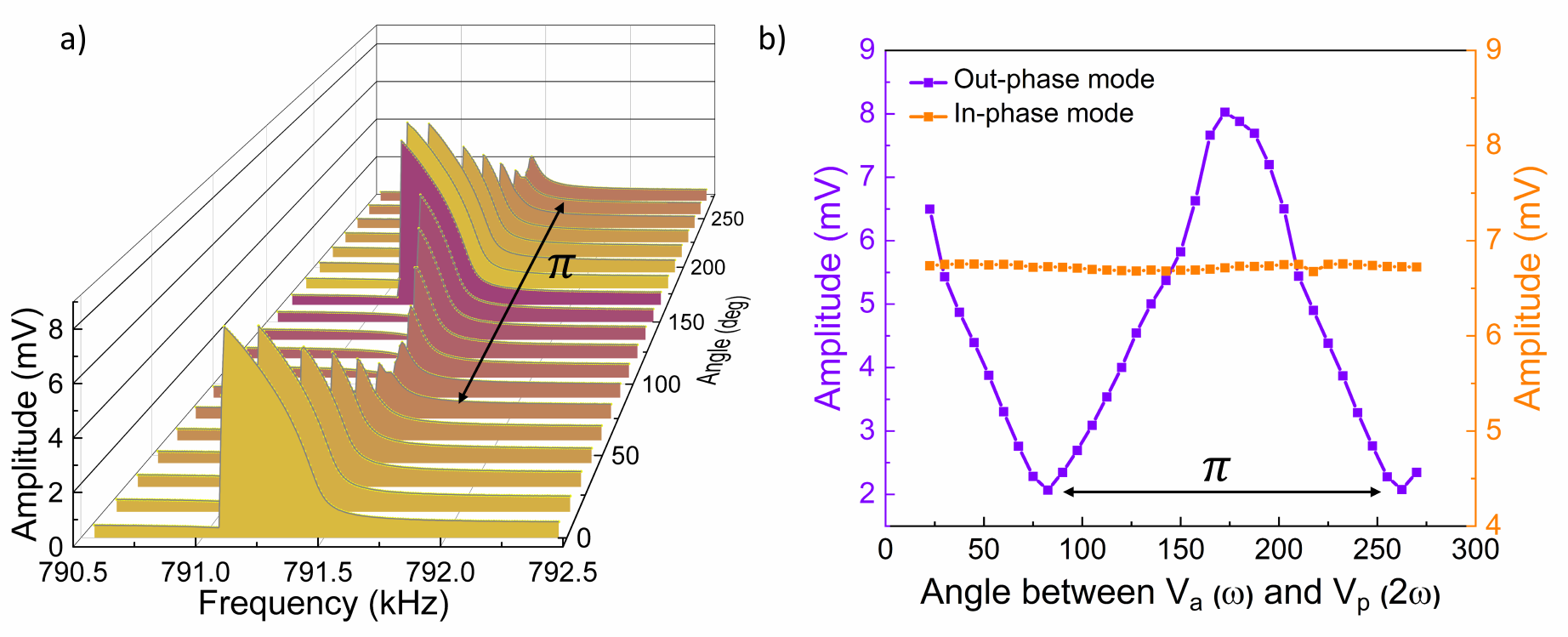}
    \caption{Device parametric with harmonic measurement.(a) Frequency response of out-phase mode with varying phase angle between direct and parametric drive (b) Variation of gain with phase for out-phase mode (blue) and in-phase mode (red).}
    \label{drive_dir_para}
\end{figure}

In addition to the parametric drive, a harmonic drive signal at the actuation electrode is applied. The parametric pump is kept below the critical pump excitation to prevent the system from being at self-oscillation. By varying the phase angle between the drive signal at $\omega$ and the pump signal at $2\omega$, the system undergoes higher amplification to lower amplification in a sinusoidal fashion \cite{Nishta_parametric}, since the gain of the parametric system depends on the relative phase angle. The in-phase mode response shows no oscillation behavior due to the negligible effect of parametric excitation. In contrast, the out-phase mode exhibits oscillation as the angle changes and the oscillation is further illustrated in \autoref{drive_dir_para} {b}.The equation of motion of the parametric oscillator for the out-phase mode can be written as:
\begin{equation}\label{eqn 4_23}
    m\Ddot{v}+c\dot{v}+(k_1-f_b)v+2k_{c1}(1+\beta sin(2\omega t))v-2k_{c3}v^3=f_a cos(\omega t+\theta)
\end{equation}

Here, $\theta$ is the phase angle between the harmonic drive and pump drive signal. 

The \autoref{drive_dir_para} {a} depicts the behavior of the nonlinear response with the phase angle of the drive and the parametric pump. The measured oscillation between maximum (amplification) and minimum (de-amplification) while varying phase angle is shown \autoref{drive_dir_para} {b}. The in-phase mode shows no change in amplitude due to the negligible impact of nonlinearity, while the out-phase mode exhibits a cosine oscillation.

The phase angle at which the maxima and minima occur is $172.5^\circ$ and $82.5^\circ$ respectively. The phase difference between the maxima and minima of the oscillation is $\pi/2$, and the successive maxima and/or minima are $\pi$. The oscillation provides insights into the system's amplification and helps identify the optimal operating point to maximize signal strength for applications such as mass sensing, logic, memory, etc \cite{Turner_mass_sensing,Villanueva_parametric_oscillator,Yamaguchi_GaAs_nanoresonator,Bagheri_memory}. The system's functionality lies in its degenerate phase modes, where one phase mode can be excited for the on-state and another for the off-state, enabling its application as a logic system. Therefore, it can be a powerful device for leveraging degenerate modes in such applications.

\subsection{Parametric Gain}

\begin{figure}[!h]
    \centering
    \includegraphics[width =\linewidth]{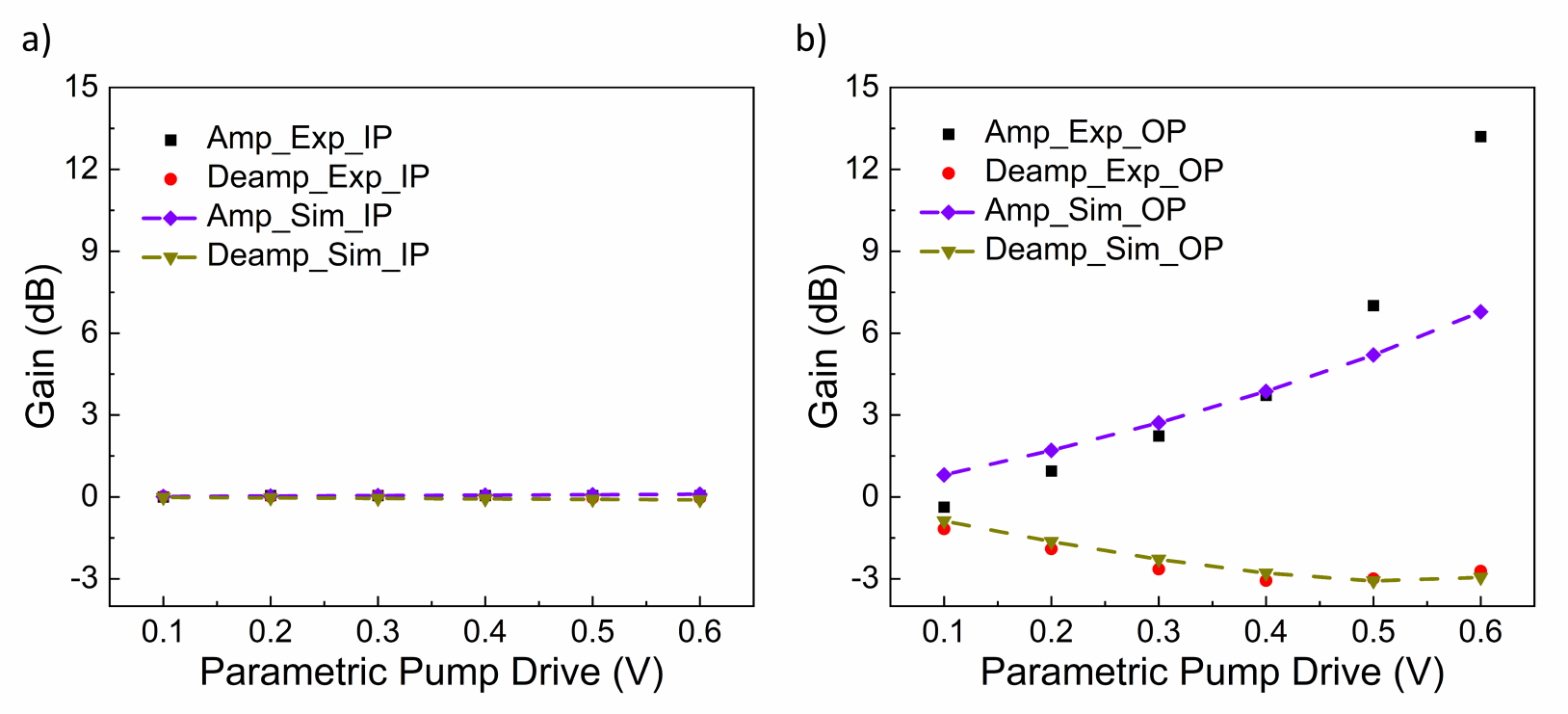}
    \caption{(a). In-phase mode gain shows no effect of parametric pump drive at the angle of maxima and minima of the parametric oscillation, whereas, (b). Out-phase mode gain shows the effect of amplification at the maxima of the parametric oscillation and suppression at the minima of the oscillation}
    \label{gain_para}
\end{figure}

The \autoref{gain_para} shows the behavior of the phase modes gain when the parametric drive is applied at the maxima ($\theta=172.5^\circ$) and minima ($\theta=82.5^\circ$) relative phase angles while keeping the harmonic drive constant. The parametric gain for out-phase mode shows amplification of $\sim 13$ dB and the parametric compression/de-amplification of $\sim 3$ dB. Further increasing the parametric pump drive pushes the system into the instability regime \cite{Gary_Steele_parametric}.

The gain is additionally simulated by determining the critical amplitude of the parametric drive at which the onset of parametric oscillations occurs. The phase-dependent modulation of amplification and de-amplification, corresponding to the maxima and minima of the relative phase between the harmonic and parametric drives, is given by\cite{Nishta_parametric}:

\begin{equation}\label{eqn 4_24}
    G=\sqrt{\dfrac{cos^2(\theta+\pi/2)}{1+(\dfrac{V_p}{V_c})^2}+\dfrac{sin^2(\theta+\pi/2)}{1-(\dfrac{V_p}{V_c})^2}}
\end{equation}
Here, $V_p$ and $V_c$ represent parametric pump drive and critical parametric drive respectively. The phase lag due to the instrument is also accounted for by adding a $\pi/2$ phase shift to the phase angle, as angle change was performed at the parametric drive demodulator of LIA \cite{Pi_2_angle_add}. The simulation shows a good match with the measurements. At higher actuation drive, higher order nonlinear coupling leads to a significantly enhanced gain in the out-phase mode, deviating from the theoretical predictions. Further investigation is required to fully understand the role and influence of higher-order coupling and is beyond the scope of this work.

\section{Conclusion}

We have demonstrated the effect of coupling in a double-ended tuning fork (DETF) resonator using degenerate modes to explore nonlinear dynamics and parametric amplification. By engineering a system of coupled resonators, we achieve selective parametric excitation, which plays a crucial role in defining the observed nonlinear behavior and gain characteristics. The out-phase mode exhibits stronger nonlinearity than the in-phase mode, consistent with the greater influence of coupling beam in its motion. Moreover, parametric amplification requiring careful tuning of the system parameters was realized exclusively in the out-phase mode. These results highlight the potential of coupling-based control for selective activation of nonlinear responses and parametric gain in coupled resonators. This work establishes a promising pathway for future applications in reconfigurable mechanical memory, logic, amplification, and sensing platforms.


\begin{acknowledgments}

We wish to acknowledge the support of the Centre for Nanoscience and Engineering, IISc, the National Nano Fabrication Centre (NNFC), and the Micro-Nano Characterization Facility (MNCF). Vishnu Kumar gratefully acknowledges the MHRD, Govt. of India, for providing us with the necessary funding and fellowship to pursue research work. Saurabh A. Chandorkar acknowledges Indian Space Research Organization, Government of India (GoI) under Grant DS 2B13012(2)/41/2018-Sec.2, by the Ministry of Electronics and Information Technology, GoI under 25(2)/2020 ESDA DT.28.05.2020 and Ministry of Human Resource and Development, GoI Grant SR/MHRD 18 0017. Saurabh A. Chandorkar also acknowledges DRDO JATP: DFTM/02/3125/M/12/MNSST-03.

\end{acknowledgments}

\bibliography{references}

\begin{thebibliography}{31}%
\makeatletter
\providecommand \@ifxundefined [1]{%
 \@ifx{#1\undefined}
}%
\providecommand \@ifnum [1]{%
 \ifnum #1\expandafter \@firstoftwo
 \else \expandafter \@secondoftwo
 \fi
}%
\providecommand \@ifx [1]{%
 \ifx #1\expandafter \@firstoftwo
 \else \expandafter \@secondoftwo
 \fi
}%
\providecommand \natexlab [1]{#1}%
\providecommand \enquote  [1]{``#1''}%
\providecommand \bibnamefont  [1]{#1}%
\providecommand \bibfnamefont [1]{#1}%
\providecommand \citenamefont [1]{#1}%
\providecommand \href@noop [0]{\@secondoftwo}%
\providecommand \href [0]{\begingroup \@sanitize@url \@href}%
\providecommand \@href[1]{\@@startlink{#1}\@@href}%
\providecommand \@@href[1]{\endgroup#1\@@endlink}%
\providecommand \@sanitize@url [0]{\catcode `\\12\catcode `\$12\catcode `\&12\catcode `\#12\catcode `\^12\catcode `\_12\catcode `\%12\relax}%
\providecommand \@@startlink[1]{}%
\providecommand \@@endlink[0]{}%
\providecommand \url  [0]{\begingroup\@sanitize@url \@url }%
\providecommand \@url [1]{\endgroup\@href {#1}{\urlprefix }}%
\providecommand \urlprefix  [0]{URL }%
\providecommand \Eprint [0]{\href }%
\providecommand \doibase [0]{https://doi.org/}%
\providecommand \selectlanguage [0]{\@gobble}%
\providecommand \bibinfo  [0]{\@secondoftwo}%
\providecommand \bibfield  [0]{\@secondoftwo}%
\providecommand \translation [1]{[#1]}%
\providecommand \BibitemOpen [0]{}%
\providecommand \bibitemStop [0]{}%
\providecommand \bibitemNoStop [0]{.\EOS\space}%
\providecommand \EOS [0]{\spacefactor3000\relax}%
\providecommand \BibitemShut  [1]{\csname bibitem#1\endcsname}%
\let\auto@bib@innerbib\@empty
\bibitem [{\citenamefont {Prasad}\ \emph {et~al.}(2017)\citenamefont {Prasad}, \citenamefont {Arora},\ and\ \citenamefont {Naik}}]{Nishta_parametric}%
  \BibitemOpen
  \bibfield  {author} {\bibinfo {author} {\bibfnamefont {P.}~\bibnamefont {Prasad}}, \bibinfo {author} {\bibfnamefont {N.}~\bibnamefont {Arora}},\ and\ \bibinfo {author} {\bibfnamefont {A.~K.}\ \bibnamefont {Naik}},\ }\bibfield  {title} {\bibinfo {title} {Parametric amplification in mos2 drum resonator},\ }\href {https://doi.org/10.1039/c7nr05721k} {\bibfield  {journal} {\bibinfo  {journal} {Nanoscale}\ }\textbf {\bibinfo {volume} {9}},\ \bibinfo {pages} {18299–18304} (\bibinfo {year} {2017})}\BibitemShut {NoStop}%
\bibitem [{\citenamefont {Zhang}\ \emph {et~al.}(2002)\citenamefont {Zhang}, \citenamefont {Baskaran},\ and\ \citenamefont {Turner}}]{Turner_mass_sensing}%
  \BibitemOpen
  \bibfield  {author} {\bibinfo {author} {\bibfnamefont {W.}~\bibnamefont {Zhang}}, \bibinfo {author} {\bibfnamefont {R.}~\bibnamefont {Baskaran}},\ and\ \bibinfo {author} {\bibfnamefont {K.~L.}\ \bibnamefont {Turner}},\ }\bibfield  {title} {\bibinfo {title} {Effect of cubic nonlinearity on auto-parametrically amplified resonant mems mass sensor},\ }\href {https://doi.org/10.1016/s0924-4247(02)00299-6} {\bibfield  {journal} {\bibinfo  {journal} {Sensors and Actuators A: Physical}\ }\textbf {\bibinfo {volume} {102}},\ \bibinfo {pages} {139–150} (\bibinfo {year} {2002})}\BibitemShut {NoStop}%
\bibitem [{\citenamefont {Tadokoro}\ and\ \citenamefont {Tanaka}(2021)}]{nonlinear_memory}%
  \BibitemOpen
  \bibfield  {author} {\bibinfo {author} {\bibfnamefont {Y.}~\bibnamefont {Tadokoro}}\ and\ \bibinfo {author} {\bibfnamefont {H.}~\bibnamefont {Tanaka}},\ }\bibfield  {title} {\bibinfo {title} {Highly sensitive implementation of logic gates with a nonlinear nanomechanical resonator},\ }\bibfield  {journal} {\bibinfo  {journal} {Physical Review Applied}\ }\textbf {\bibinfo {volume} {15}},\ \href {https://doi.org/10.1103/physrevapplied.15.024058} {10.1103/physrevapplied.15.024058} (\bibinfo {year} {2021})\BibitemShut {NoStop}%
\bibitem [{\citenamefont {Karabalin}\ \emph {et~al.}(2011)\citenamefont {Karabalin}, \citenamefont {Lifshitz}, \citenamefont {Cross}, \citenamefont {Matheny}, \citenamefont {Masmanidis},\ and\ \citenamefont {Roukes}}]{Topo_amplifier}%
  \BibitemOpen
  \bibfield  {author} {\bibinfo {author} {\bibfnamefont {R.~B.}\ \bibnamefont {Karabalin}}, \bibinfo {author} {\bibfnamefont {R.}~\bibnamefont {Lifshitz}}, \bibinfo {author} {\bibfnamefont {M.~C.}\ \bibnamefont {Cross}}, \bibinfo {author} {\bibfnamefont {M.~H.}\ \bibnamefont {Matheny}}, \bibinfo {author} {\bibfnamefont {S.~C.}\ \bibnamefont {Masmanidis}},\ and\ \bibinfo {author} {\bibfnamefont {M.~L.}\ \bibnamefont {Roukes}},\ }\bibfield  {title} {\bibinfo {title} {Signal amplification by sensitive control of bifurcation topology},\ }\bibfield  {journal} {\bibinfo  {journal} {Physical Review Letters}\ }\textbf {\bibinfo {volume} {106}},\ \href {https://doi.org/10.1103/physrevlett.106.094102} {10.1103/physrevlett.106.094102} (\bibinfo {year} {2011})\BibitemShut {NoStop}%
\bibitem [{\citenamefont {Szorkovszky}\ \emph {et~al.}(2011{\natexlab{a}})\citenamefont {Szorkovszky}, \citenamefont {Doherty}, \citenamefont {Harris},\ and\ \citenamefont {Bowen}}]{Bowen_squeezing_quantum_sensing}%
  \BibitemOpen
  \bibfield  {author} {\bibinfo {author} {\bibfnamefont {A.}~\bibnamefont {Szorkovszky}}, \bibinfo {author} {\bibfnamefont {A.~C.}\ \bibnamefont {Doherty}}, \bibinfo {author} {\bibfnamefont {G.~I.}\ \bibnamefont {Harris}},\ and\ \bibinfo {author} {\bibfnamefont {W.~P.}\ \bibnamefont {Bowen}},\ }\bibfield  {title} {\bibinfo {title} {Mechanical squeezing via parametric amplification and weak measurement},\ }\bibfield  {journal} {\bibinfo  {journal} {Physical Review Letters}\ }\textbf {\bibinfo {volume} {107}},\ \href {https://doi.org/10.1103/physrevlett.107.213603} {10.1103/physrevlett.107.213603} (\bibinfo {year} {2011}{\natexlab{a}})\BibitemShut {NoStop}%
\bibitem [{\citenamefont {Zhang}\ \emph {et~al.}(2024)\citenamefont {Zhang}, \citenamefont {Wang}, \citenamefont {Zhang}, \citenamefont {Jiao}, \citenamefont {Zuo}, \citenamefont {\"{O}zdemir}, \citenamefont {Qiu}, \citenamefont {Nori},\ and\ \citenamefont {Jing}}]{quantum_sensing}%
  \BibitemOpen
  \bibfield  {author} {\bibinfo {author} {\bibfnamefont {S.-D.}\ \bibnamefont {Zhang}}, \bibinfo {author} {\bibfnamefont {J.}~\bibnamefont {Wang}}, \bibinfo {author} {\bibfnamefont {Q.}~\bibnamefont {Zhang}}, \bibinfo {author} {\bibfnamefont {Y.-F.}\ \bibnamefont {Jiao}}, \bibinfo {author} {\bibfnamefont {Y.-L.}\ \bibnamefont {Zuo}}, \bibinfo {author} {\bibfnamefont {Å.~K.}\ \bibnamefont {\"{O}zdemir}}, \bibinfo {author} {\bibfnamefont {C.-W.}\ \bibnamefont {Qiu}}, \bibinfo {author} {\bibfnamefont {F.}~\bibnamefont {Nori}},\ and\ \bibinfo {author} {\bibfnamefont {H.}~\bibnamefont {Jing}},\ }\bibfield  {title} {\bibinfo {title} {Squeezing-enhanced quantum sensing with quadratic optomechanics},\ }\href {https://doi.org/10.1364/opticaq.523480} {\bibfield  {journal} {\bibinfo  {journal} {Optica Quantum}\ }\textbf {\bibinfo {volume} {2}},\ \bibinfo {pages} {222} (\bibinfo {year} {2024})}\BibitemShut {NoStop}%
\bibitem [{\citenamefont {Guo}\ \emph {et~al.}(2023)\citenamefont {Guo}, \citenamefont {Fang}, \citenamefont {Chen}, \citenamefont {Li}, \citenamefont {Chen}, \citenamefont {Zhou}, \citenamefont {Wang}, \citenamefont {Song}, \citenamefont {Arutyunov}, \citenamefont {Guo}, \citenamefont {Wang},\ and\ \citenamefont {Deng}}]{mode_coupling_review}%
  \BibitemOpen
  \bibfield  {author} {\bibinfo {author} {\bibfnamefont {M.}~\bibnamefont {Guo}}, \bibinfo {author} {\bibfnamefont {J.}~\bibnamefont {Fang}}, \bibinfo {author} {\bibfnamefont {J.}~\bibnamefont {Chen}}, \bibinfo {author} {\bibfnamefont {B.}~\bibnamefont {Li}}, \bibinfo {author} {\bibfnamefont {H.}~\bibnamefont {Chen}}, \bibinfo {author} {\bibfnamefont {Q.}~\bibnamefont {Zhou}}, \bibinfo {author} {\bibfnamefont {Y.}~\bibnamefont {Wang}}, \bibinfo {author} {\bibfnamefont {H.}~\bibnamefont {Song}}, \bibinfo {author} {\bibfnamefont {K.~Y.}\ \bibnamefont {Arutyunov}}, \bibinfo {author} {\bibfnamefont {G.}~\bibnamefont {Guo}}, \bibinfo {author} {\bibfnamefont {Z.}~\bibnamefont {Wang}},\ and\ \bibinfo {author} {\bibfnamefont {G.}~\bibnamefont {Deng}},\ }\bibfield  {title} {\bibinfo {title} {Mode coupling in electromechanical systems: Recent advances and applications},\ }\bibfield  {journal} {\bibinfo  {journal} {Advanced Electronic Materials}\ }\textbf {\bibinfo {volume} {9}},\ \href
  {https://doi.org/10.1002/aelm.202201305} {10.1002/aelm.202201305} (\bibinfo {year} {2023})\BibitemShut {NoStop}%
\bibitem [{\citenamefont {Prasad}\ \emph {et~al.}(2019)\citenamefont {Prasad}, \citenamefont {Arora},\ and\ \citenamefont {Naik}}]{Nishta_coupling}%
  \BibitemOpen
  \bibfield  {author} {\bibinfo {author} {\bibfnamefont {P.}~\bibnamefont {Prasad}}, \bibinfo {author} {\bibfnamefont {N.}~\bibnamefont {Arora}},\ and\ \bibinfo {author} {\bibfnamefont {A.~K.}\ \bibnamefont {Naik}},\ }\bibfield  {title} {\bibinfo {title} {Gate tunable cooperativity between vibrational modes},\ }\href {https://doi.org/10.1021/acs.nanolett.9b01219} {\bibfield  {journal} {\bibinfo  {journal} {Nano Letters}\ }\textbf {\bibinfo {volume} {19}},\ \bibinfo {pages} {5862–5867} (\bibinfo {year} {2019})}\BibitemShut {NoStop}%
\bibitem [{\citenamefont {Xu}\ \emph {et~al.}(2024)\citenamefont {Xu}, \citenamefont {Yang}, \citenamefont {Song},\ and\ \citenamefont {Wei}}]{high_sensitivity}%
  \BibitemOpen
  \bibfield  {author} {\bibinfo {author} {\bibfnamefont {Y.}~\bibnamefont {Xu}}, \bibinfo {author} {\bibfnamefont {Q.}~\bibnamefont {Yang}}, \bibinfo {author} {\bibfnamefont {J.}~\bibnamefont {Song}},\ and\ \bibinfo {author} {\bibfnamefont {X.}~\bibnamefont {Wei}},\ }\bibfield  {title} {\bibinfo {title} {Sensitivity enhancement of nonlinear micromechanical sensors using parametric symmetry breaking},\ }\bibfield  {journal} {\bibinfo  {journal} {Microsystems and Nanoengineering}\ }\textbf {\bibinfo {volume} {10}},\ \href {https://doi.org/10.1038/s41378-024-00784-4} {10.1038/s41378-024-00784-4} (\bibinfo {year} {2024})\BibitemShut {NoStop}%
\bibitem [{\citenamefont {Szorkovszky}\ \emph {et~al.}(2011{\natexlab{b}})\citenamefont {Szorkovszky}, \citenamefont {Doherty}, \citenamefont {Harris},\ and\ \citenamefont {Bowen}}]{Bowen_para_cooling}%
  \BibitemOpen
  \bibfield  {author} {\bibinfo {author} {\bibfnamefont {A.}~\bibnamefont {Szorkovszky}}, \bibinfo {author} {\bibfnamefont {A.~C.}\ \bibnamefont {Doherty}}, \bibinfo {author} {\bibfnamefont {G.~I.}\ \bibnamefont {Harris}},\ and\ \bibinfo {author} {\bibfnamefont {W.~P.}\ \bibnamefont {Bowen}},\ }\bibfield  {title} {\bibinfo {title} {Mechanical squeezing via parametric amplification and weak measurement},\ }\bibfield  {journal} {\bibinfo  {journal} {Physical Review Letters}\ }\textbf {\bibinfo {volume} {107}},\ \href {https://doi.org/10.1103/physrevlett.107.213603} {10.1103/physrevlett.107.213603} (\bibinfo {year} {2011}{\natexlab{b}})\BibitemShut {NoStop}%
\bibitem [{\citenamefont {Thiruvenkatanathan}\ \emph {et~al.}(2010)\citenamefont {Thiruvenkatanathan}, \citenamefont {Yan}, \citenamefont {Woodhouse}, \citenamefont {Aziz},\ and\ \citenamefont {Seshia}}]{Seshia_para_sensing}%
  \BibitemOpen
  \bibfield  {author} {\bibinfo {author} {\bibfnamefont {P.}~\bibnamefont {Thiruvenkatanathan}}, \bibinfo {author} {\bibfnamefont {J.}~\bibnamefont {Yan}}, \bibinfo {author} {\bibfnamefont {J.}~\bibnamefont {Woodhouse}}, \bibinfo {author} {\bibfnamefont {A.}~\bibnamefont {Aziz}},\ and\ \bibinfo {author} {\bibfnamefont {A.~A.}\ \bibnamefont {Seshia}},\ }\bibfield  {title} {\bibinfo {title} {Ultrasensitive mode-localized mass sensor with electrically tunable parametric sensitivity},\ }\bibfield  {journal} {\bibinfo  {journal} {Applied Physics Letters}\ }\textbf {\bibinfo {volume} {96}},\ \href {https://doi.org/10.1063/1.3315877} {10.1063/1.3315877} (\bibinfo {year} {2010})\BibitemShut {NoStop}%
\bibitem [{\citenamefont {Miller}\ \emph {et~al.}(2021)\citenamefont {Miller}, \citenamefont {Gomez-Franco}, \citenamefont {Shin}, \citenamefont {Kwon},\ and\ \citenamefont {Kenny}}]{Frequency_stability}%
  \BibitemOpen
  \bibfield  {author} {\bibinfo {author} {\bibfnamefont {J.~M.~L.}\ \bibnamefont {Miller}}, \bibinfo {author} {\bibfnamefont {A.}~\bibnamefont {Gomez-Franco}}, \bibinfo {author} {\bibfnamefont {D.~D.}\ \bibnamefont {Shin}}, \bibinfo {author} {\bibfnamefont {H.-K.}\ \bibnamefont {Kwon}},\ and\ \bibinfo {author} {\bibfnamefont {T.~W.}\ \bibnamefont {Kenny}},\ }\bibfield  {title} {\bibinfo {title} {Amplitude stabilization of micromechanical oscillators using engineered nonlinearity},\ }\bibfield  {journal} {\bibinfo  {journal} {Physical Review Research}\ }\textbf {\bibinfo {volume} {3}},\ \href {https://doi.org/10.1103/physrevresearch.3.033268} {10.1103/physrevresearch.3.033268} (\bibinfo {year} {2021})\BibitemShut {NoStop}%
\bibitem [{\citenamefont {Okamoto}\ \emph {et~al.}(2013)\citenamefont {Okamoto}, \citenamefont {Gourgout}, \citenamefont {Chang}, \citenamefont {Onomitsu}, \citenamefont {Mahboob}, \citenamefont {Chang},\ and\ \citenamefont {Yamaguchi}}]{Yamaguchi_coherentphonon}%
  \BibitemOpen
  \bibfield  {author} {\bibinfo {author} {\bibfnamefont {H.}~\bibnamefont {Okamoto}}, \bibinfo {author} {\bibfnamefont {A.}~\bibnamefont {Gourgout}}, \bibinfo {author} {\bibfnamefont {C.-Y.}\ \bibnamefont {Chang}}, \bibinfo {author} {\bibfnamefont {K.}~\bibnamefont {Onomitsu}}, \bibinfo {author} {\bibfnamefont {I.}~\bibnamefont {Mahboob}}, \bibinfo {author} {\bibfnamefont {E.~Y.}\ \bibnamefont {Chang}},\ and\ \bibinfo {author} {\bibfnamefont {H.}~\bibnamefont {Yamaguchi}},\ }\bibfield  {title} {\bibinfo {title} {Coherent phonon manipulation in coupled mechanical resonators},\ }\href {https://doi.org/10.1038/nphys2665} {\bibfield  {journal} {\bibinfo  {journal} {Nature Physics}\ }\textbf {\bibinfo {volume} {9}},\ \bibinfo {pages} {480–484} (\bibinfo {year} {2013})}\BibitemShut {NoStop}%
\bibitem [{\citenamefont {Zhang}\ \emph {et~al.}(2025)\citenamefont {Zhang}, \citenamefont {Li}, \citenamefont {Sun}, \citenamefont {Kirkbride}, \citenamefont {Teng}, \citenamefont {Liu}, \citenamefont {Chen}, \citenamefont {Parajuli}, \citenamefont {Pandit}, \citenamefont {Sobreviela}, \citenamefont {Zhao}, \citenamefont {Yuan}, \citenamefont {Chang},\ and\ \citenamefont {Seshia}}]{Seshia_coupled_nonlinear}%
  \BibitemOpen
  \bibfield  {author} {\bibinfo {author} {\bibfnamefont {H.}~\bibnamefont {Zhang}}, \bibinfo {author} {\bibfnamefont {H.}~\bibnamefont {Li}}, \bibinfo {author} {\bibfnamefont {J.}~\bibnamefont {Sun}}, \bibinfo {author} {\bibfnamefont {S.}~\bibnamefont {Kirkbride}}, \bibinfo {author} {\bibfnamefont {G.}~\bibnamefont {Teng}}, \bibinfo {author} {\bibfnamefont {Z.}~\bibnamefont {Liu}}, \bibinfo {author} {\bibfnamefont {D.}~\bibnamefont {Chen}}, \bibinfo {author} {\bibfnamefont {M.}~\bibnamefont {Parajuli}}, \bibinfo {author} {\bibfnamefont {M.}~\bibnamefont {Pandit}}, \bibinfo {author} {\bibfnamefont {G.}~\bibnamefont {Sobreviela}}, \bibinfo {author} {\bibfnamefont {C.}~\bibnamefont {Zhao}}, \bibinfo {author} {\bibfnamefont {W.}~\bibnamefont {Yuan}}, \bibinfo {author} {\bibfnamefont {H.}~\bibnamefont {Chang}},\ and\ \bibinfo {author} {\bibfnamefont {A.~A.}\ \bibnamefont {Seshia}},\ }\bibfield  {title} {\bibinfo {title} {Coherent energy transfer in coupled nonlinear microelectromechanical resonators},\ }\bibfield
  {journal} {\bibinfo  {journal} {Nature Communications}\ }\textbf {\bibinfo {volume} {16}},\ \href {https://doi.org/10.1038/s41467-025-59292-2} {10.1038/s41467-025-59292-2} (\bibinfo {year} {2025})\BibitemShut {NoStop}%
\bibitem [{\citenamefont {Matheny}\ \emph {et~al.}(2014)\citenamefont {Matheny}, \citenamefont {Grau}, \citenamefont {Villanueva}, \citenamefont {Karabalin}, \citenamefont {Cross},\ and\ \citenamefont {Roukes}}]{synchronization}%
  \BibitemOpen
  \bibfield  {author} {\bibinfo {author} {\bibfnamefont {M.~H.}\ \bibnamefont {Matheny}}, \bibinfo {author} {\bibfnamefont {M.}~\bibnamefont {Grau}}, \bibinfo {author} {\bibfnamefont {L.~G.}\ \bibnamefont {Villanueva}}, \bibinfo {author} {\bibfnamefont {R.~B.}\ \bibnamefont {Karabalin}}, \bibinfo {author} {\bibfnamefont {M.}~\bibnamefont {Cross}},\ and\ \bibinfo {author} {\bibfnamefont {M.~L.}\ \bibnamefont {Roukes}},\ }\bibfield  {title} {\bibinfo {title} {Phase synchronization of two anharmonic nanomechanical oscillators},\ }\bibfield  {journal} {\bibinfo  {journal} {Physical Review Letters}\ }\textbf {\bibinfo {volume} {112}},\ \href {https://doi.org/10.1103/physrevlett.112.014101} {10.1103/physrevlett.112.014101} (\bibinfo {year} {2014})\BibitemShut {NoStop}%
\bibitem [{\citenamefont {Doster}\ \emph {et~al.}(2019)\citenamefont {Doster}, \citenamefont {Hoenl}, \citenamefont {Lorenz}, \citenamefont {Paulitschke},\ and\ \citenamefont {Weig}}]{mode_hybridization}%
  \BibitemOpen
  \bibfield  {author} {\bibinfo {author} {\bibfnamefont {J.}~\bibnamefont {Doster}}, \bibinfo {author} {\bibfnamefont {S.}~\bibnamefont {Hoenl}}, \bibinfo {author} {\bibfnamefont {H.}~\bibnamefont {Lorenz}}, \bibinfo {author} {\bibfnamefont {P.}~\bibnamefont {Paulitschke}},\ and\ \bibinfo {author} {\bibfnamefont {E.~M.}\ \bibnamefont {Weig}},\ }\bibfield  {title} {\bibinfo {title} {Collective dynamics of strain-coupled nanomechanical pillar resonators},\ }\bibfield  {journal} {\bibinfo  {journal} {Nature Communications}\ }\textbf {\bibinfo {volume} {10}},\ \href {https://doi.org/10.1038/s41467-019-13309-9} {10.1038/s41467-019-13309-9} (\bibinfo {year} {2019})\BibitemShut {NoStop}%
\bibitem [{\citenamefont {Gil-Santos}\ \emph {et~al.}(2009)\citenamefont {Gil-Santos}, \citenamefont {Ramos}, \citenamefont {Jana}, \citenamefont {Calleja}, \citenamefont {Raman},\ and\ \citenamefont {Tamayo}}]{sensing_nano}%
  \BibitemOpen
  \bibfield  {author} {\bibinfo {author} {\bibfnamefont {E.}~\bibnamefont {Gil-Santos}}, \bibinfo {author} {\bibfnamefont {D.}~\bibnamefont {Ramos}}, \bibinfo {author} {\bibfnamefont {A.}~\bibnamefont {Jana}}, \bibinfo {author} {\bibfnamefont {M.}~\bibnamefont {Calleja}}, \bibinfo {author} {\bibfnamefont {A.}~\bibnamefont {Raman}},\ and\ \bibinfo {author} {\bibfnamefont {J.}~\bibnamefont {Tamayo}},\ }\bibfield  {title} {\bibinfo {title} {Mass sensing based on deterministic and stochastic responses of elastically coupled nanocantilevers},\ }\href {https://doi.org/10.1021/nl902350b} {\bibfield  {journal} {\bibinfo  {journal} {Nano Letters}\ }\textbf {\bibinfo {volume} {9}},\ \bibinfo {pages} {4122–4127} (\bibinfo {year} {2009})}\BibitemShut {NoStop}%
\bibitem [{\citenamefont {Han}\ \emph {et~al.}(2024)\citenamefont {Han}, \citenamefont {Wang}, \citenamefont {Zhang}, \citenamefont {Dykman},\ and\ \citenamefont {Chan}}]{Chan_Ising}%
  \BibitemOpen
  \bibfield  {author} {\bibinfo {author} {\bibfnamefont {C.}~\bibnamefont {Han}}, \bibinfo {author} {\bibfnamefont {M.}~\bibnamefont {Wang}}, \bibinfo {author} {\bibfnamefont {B.}~\bibnamefont {Zhang}}, \bibinfo {author} {\bibfnamefont {M.~I.}\ \bibnamefont {Dykman}},\ and\ \bibinfo {author} {\bibfnamefont {H.~B.}\ \bibnamefont {Chan}},\ }\bibfield  {title} {\bibinfo {title} {Coupled parametric oscillators: From disorder-induced current to asymmetric ising model},\ }\bibfield  {journal} {\bibinfo  {journal} {Physical Review Research}\ }\textbf {\bibinfo {volume} {6}},\ \href {https://doi.org/10.1103/physrevresearch.6.023162} {10.1103/physrevresearch.6.023162} (\bibinfo {year} {2024})\BibitemShut {NoStop}%
\bibitem [{\citenamefont {Zhou}\ \emph {et~al.}(2019)\citenamefont {Zhou}, \citenamefont {Zhao}, \citenamefont {Xiao}, \citenamefont {Sun}, \citenamefont {Sobreviela}, \citenamefont {Gerrard}, \citenamefont {Chen}, \citenamefont {Flader}, \citenamefont {Kenny}, \citenamefont {Wu},\ and\ \citenamefont {Seshia}}]{Seshia_electrostat_modecoupling}%
  \BibitemOpen
  \bibfield  {author} {\bibinfo {author} {\bibfnamefont {X.}~\bibnamefont {Zhou}}, \bibinfo {author} {\bibfnamefont {C.}~\bibnamefont {Zhao}}, \bibinfo {author} {\bibfnamefont {D.}~\bibnamefont {Xiao}}, \bibinfo {author} {\bibfnamefont {J.}~\bibnamefont {Sun}}, \bibinfo {author} {\bibfnamefont {G.}~\bibnamefont {Sobreviela}}, \bibinfo {author} {\bibfnamefont {D.~D.}\ \bibnamefont {Gerrard}}, \bibinfo {author} {\bibfnamefont {Y.}~\bibnamefont {Chen}}, \bibinfo {author} {\bibfnamefont {I.}~\bibnamefont {Flader}}, \bibinfo {author} {\bibfnamefont {T.~W.}\ \bibnamefont {Kenny}}, \bibinfo {author} {\bibfnamefont {X.}~\bibnamefont {Wu}},\ and\ \bibinfo {author} {\bibfnamefont {A.~A.}\ \bibnamefont {Seshia}},\ }\bibfield  {title} {\bibinfo {title} {Dynamic modulation of modal coupling in microelectromechanical gyroscopic ring resonators},\ }\bibfield  {journal} {\bibinfo  {journal} {Nature Communications}\ }\textbf {\bibinfo {volume} {10}},\ \href {https://doi.org/10.1038/s41467-019-12796-0}
  {10.1038/s41467-019-12796-0} (\bibinfo {year} {2019})\BibitemShut {NoStop}%
\bibitem [{\citenamefont {Hälg}\ \emph {et~al.}(2022)\citenamefont {Hälg}, \citenamefont {Gisler}, \citenamefont {Langman}, \citenamefont {Misra}, \citenamefont {Zilberberg}, \citenamefont {Schliesser}, \citenamefont {Degen},\ and\ \citenamefont {Eichler}}]{Halg_sin_parametric}%
  \BibitemOpen
  \bibfield  {author} {\bibinfo {author} {\bibfnamefont {D.}~\bibnamefont {Hälg}}, \bibinfo {author} {\bibfnamefont {T.}~\bibnamefont {Gisler}}, \bibinfo {author} {\bibfnamefont {E.~C.}\ \bibnamefont {Langman}}, \bibinfo {author} {\bibfnamefont {S.}~\bibnamefont {Misra}}, \bibinfo {author} {\bibfnamefont {O.}~\bibnamefont {Zilberberg}}, \bibinfo {author} {\bibfnamefont {A.}~\bibnamefont {Schliesser}}, \bibinfo {author} {\bibfnamefont {C.~L.}\ \bibnamefont {Degen}},\ and\ \bibinfo {author} {\bibfnamefont {A.}~\bibnamefont {Eichler}},\ }\bibfield  {title} {\bibinfo {title} {Strong parametric coupling between two ultracoherent membrane modes},\ }\bibfield  {journal} {\bibinfo  {journal} {Physical Review Letters}\ }\textbf {\bibinfo {volume} {128}},\ \href {https://doi.org/10.1103/physrevlett.128.094301} {10.1103/physrevlett.128.094301} (\bibinfo {year} {2022})\BibitemShut {NoStop}%
\bibitem [{\citenamefont {Bagheri}\ \emph {et~al.}(2011)\citenamefont {Bagheri}, \citenamefont {Poot}, \citenamefont {Li}, \citenamefont {Pernice},\ and\ \citenamefont {Tang}}]{Bagheri_memory}%
  \BibitemOpen
  \bibfield  {author} {\bibinfo {author} {\bibfnamefont {M.}~\bibnamefont {Bagheri}}, \bibinfo {author} {\bibfnamefont {M.}~\bibnamefont {Poot}}, \bibinfo {author} {\bibfnamefont {M.}~\bibnamefont {Li}}, \bibinfo {author} {\bibfnamefont {W.~P.~H.}\ \bibnamefont {Pernice}},\ and\ \bibinfo {author} {\bibfnamefont {H.~X.}\ \bibnamefont {Tang}},\ }\bibfield  {title} {\bibinfo {title} {Dynamic manipulation of nanomechanical resonators in the high-amplitude regime and non-volatile mechanical memory operation},\ }\href {https://doi.org/10.1038/nnano.2011.180} {\bibfield  {journal} {\bibinfo  {journal} {Nature Nanotechnology}\ }\textbf {\bibinfo {volume} {6}},\ \bibinfo {pages} {726–732} (\bibinfo {year} {2011})}\BibitemShut {NoStop}%
\bibitem [{\citenamefont {Miao}\ \emph {et~al.}(2022)\citenamefont {Miao}, \citenamefont {Zhou}, \citenamefont {Wu}, \citenamefont {Li}, \citenamefont {Hou}, \citenamefont {Hu}, \citenamefont {Wang},\ and\ \citenamefont {Xiao}}]{Xiao_phononcavity}%
  \BibitemOpen
  \bibfield  {author} {\bibinfo {author} {\bibfnamefont {T.}~\bibnamefont {Miao}}, \bibinfo {author} {\bibfnamefont {X.}~\bibnamefont {Zhou}}, \bibinfo {author} {\bibfnamefont {X.}~\bibnamefont {Wu}}, \bibinfo {author} {\bibfnamefont {Q.}~\bibnamefont {Li}}, \bibinfo {author} {\bibfnamefont {Z.}~\bibnamefont {Hou}}, \bibinfo {author} {\bibfnamefont {X.}~\bibnamefont {Hu}}, \bibinfo {author} {\bibfnamefont {Z.}~\bibnamefont {Wang}},\ and\ \bibinfo {author} {\bibfnamefont {D.}~\bibnamefont {Xiao}},\ }\bibfield  {title} {\bibinfo {title} {Nonlinearity-mediated digitization and amplification in electromechanical phonon-cavity systems},\ }\bibfield  {journal} {\bibinfo  {journal} {Nature Communications}\ }\textbf {\bibinfo {volume} {13}},\ \href {https://doi.org/10.1038/s41467-022-29995-x} {10.1038/s41467-022-29995-x} (\bibinfo {year} {2022})\BibitemShut {NoStop}%
\bibitem [{\citenamefont {Villanueva}\ \emph {et~al.}(2011)\citenamefont {Villanueva}, \citenamefont {Karabalin}, \citenamefont {Matheny}, \citenamefont {Kenig}, \citenamefont {Cross},\ and\ \citenamefont {Roukes}}]{Villanueva_parametric_oscillator}%
  \BibitemOpen
  \bibfield  {author} {\bibinfo {author} {\bibfnamefont {L.~G.}\ \bibnamefont {Villanueva}}, \bibinfo {author} {\bibfnamefont {R.~B.}\ \bibnamefont {Karabalin}}, \bibinfo {author} {\bibfnamefont {M.~H.}\ \bibnamefont {Matheny}}, \bibinfo {author} {\bibfnamefont {E.}~\bibnamefont {Kenig}}, \bibinfo {author} {\bibfnamefont {M.~C.}\ \bibnamefont {Cross}},\ and\ \bibinfo {author} {\bibfnamefont {M.~L.}\ \bibnamefont {Roukes}},\ }\bibfield  {title} {\bibinfo {title} {A nanoscale parametric feedback oscillator},\ }\href {https://doi.org/10.1021/nl2031162} {\bibfield  {journal} {\bibinfo  {journal} {Nano Letters}\ }\textbf {\bibinfo {volume} {11}},\ \bibinfo {pages} {5054–5059} (\bibinfo {year} {2011})}\BibitemShut {NoStop}%
\bibitem [{\citenamefont {Samanta}\ \emph {et~al.}(2023)\citenamefont {Samanta}, \citenamefont {De~Bonis}, \citenamefont {Møller}, \citenamefont {Tormo-Queralt}, \citenamefont {Yang}, \citenamefont {Urgell}, \citenamefont {Stamenic}, \citenamefont {Thibeault}, \citenamefont {Jin}, \citenamefont {Czaplewski}, \citenamefont {Pistolesi},\ and\ \citenamefont {Bachtold}}]{Samanta2023}%
  \BibitemOpen
  \bibfield  {author} {\bibinfo {author} {\bibfnamefont {C.}~\bibnamefont {Samanta}}, \bibinfo {author} {\bibfnamefont {S.~L.}\ \bibnamefont {De~Bonis}}, \bibinfo {author} {\bibfnamefont {C.~B.}\ \bibnamefont {Møller}}, \bibinfo {author} {\bibfnamefont {R.}~\bibnamefont {Tormo-Queralt}}, \bibinfo {author} {\bibfnamefont {W.}~\bibnamefont {Yang}}, \bibinfo {author} {\bibfnamefont {C.}~\bibnamefont {Urgell}}, \bibinfo {author} {\bibfnamefont {B.}~\bibnamefont {Stamenic}}, \bibinfo {author} {\bibfnamefont {B.}~\bibnamefont {Thibeault}}, \bibinfo {author} {\bibfnamefont {Y.}~\bibnamefont {Jin}}, \bibinfo {author} {\bibfnamefont {D.~A.}\ \bibnamefont {Czaplewski}}, \bibinfo {author} {\bibfnamefont {F.}~\bibnamefont {Pistolesi}},\ and\ \bibinfo {author} {\bibfnamefont {A.}~\bibnamefont {Bachtold}},\ }\bibfield  {title} {\bibinfo {title} {Nonlinear nanomechanical resonators approaching the quantum ground state},\ }\href {https://doi.org/10.1038/s41567-023-02065-9} {\bibfield  {journal} {\bibinfo  {journal} {Nature
  Physics}\ }\textbf {\bibinfo {volume} {19}},\ \bibinfo {pages} {1340–1344} (\bibinfo {year} {2023})}\BibitemShut {NoStop}%
\bibitem [{\citenamefont {Arora}\ and\ \citenamefont {Naik}(2022)}]{Nishta_splitting}%
  \BibitemOpen
  \bibfield  {author} {\bibinfo {author} {\bibfnamefont {N.}~\bibnamefont {Arora}}\ and\ \bibinfo {author} {\bibfnamefont {A.~K.}\ \bibnamefont {Naik}},\ }\bibfield  {title} {\bibinfo {title} {Qualitative effect of internal resonance on the dynamics of two-dimensional resonator},\ }\href {https://doi.org/10.1088/1361-6463/ac5bc8} {\bibfield  {journal} {\bibinfo  {journal} {Journal of Physics D: Applied Physics}\ }\textbf {\bibinfo {volume} {55}},\ \bibinfo {pages} {265301} (\bibinfo {year} {2022})}\BibitemShut {NoStop}%
\bibitem [{\citenamefont {Keşkekler}\ \emph {et~al.}(2021)\citenamefont {Keşkekler}, \citenamefont {Shoshani}, \citenamefont {Lee}, \citenamefont {van~der Zant}, \citenamefont {Steeneken},\ and\ \citenamefont {Alijani}}]{Farbod_nonlinear}%
  \BibitemOpen
  \bibfield  {author} {\bibinfo {author} {\bibfnamefont {A.}~\bibnamefont {Keşkekler}}, \bibinfo {author} {\bibfnamefont {O.}~\bibnamefont {Shoshani}}, \bibinfo {author} {\bibfnamefont {M.}~\bibnamefont {Lee}}, \bibinfo {author} {\bibfnamefont {H.~S.~J.}\ \bibnamefont {van~der Zant}}, \bibinfo {author} {\bibfnamefont {P.~G.}\ \bibnamefont {Steeneken}},\ and\ \bibinfo {author} {\bibfnamefont {F.}~\bibnamefont {Alijani}},\ }\bibfield  {title} {\bibinfo {title} {Tuning nonlinear damping in graphene nanoresonators by parametric–direct internal resonance},\ }\bibfield  {journal} {\bibinfo  {journal} {Nature Communications}\ }\textbf {\bibinfo {volume} {12}},\ \href {https://doi.org/10.1038/s41467-021-21334-w} {10.1038/s41467-021-21334-w} (\bibinfo {year} {2021})\BibitemShut {NoStop}%
\bibitem [{\citenamefont {Welte}\ \emph {et~al.}(2013)\citenamefont {Welte}, \citenamefont {Kniffka},\ and\ \citenamefont {Ecker}}]{Welte_parametric}%
  \BibitemOpen
  \bibfield  {author} {\bibinfo {author} {\bibfnamefont {J.}~\bibnamefont {Welte}}, \bibinfo {author} {\bibfnamefont {T.~J.}\ \bibnamefont {Kniffka}},\ and\ \bibinfo {author} {\bibfnamefont {H.}~\bibnamefont {Ecker}},\ }\bibfield  {title} {\bibinfo {title} {Parametric excitation in a two degree of freedom mems system},\ }\href {https://doi.org/10.1155/2013/502109} {\bibfield  {journal} {\bibinfo  {journal} {Shock and Vibration}\ }\textbf {\bibinfo {volume} {20}},\ \bibinfo {pages} {1113–1124} (\bibinfo {year} {2013})}\BibitemShut {NoStop}%
\bibitem [{\citenamefont {Dobrindt}\ \emph {et~al.}(2008)\citenamefont {Dobrindt}, \citenamefont {Wilson-Rae},\ and\ \citenamefont {Kippenberg}}]{Dobrindt_parametric}%
  \BibitemOpen
  \bibfield  {author} {\bibinfo {author} {\bibfnamefont {J.~M.}\ \bibnamefont {Dobrindt}}, \bibinfo {author} {\bibfnamefont {I.}~\bibnamefont {Wilson-Rae}},\ and\ \bibinfo {author} {\bibfnamefont {T.~J.}\ \bibnamefont {Kippenberg}},\ }\bibfield  {title} {\bibinfo {title} {Parametric normal-mode splitting in cavity optomechanics},\ }\bibfield  {journal} {\bibinfo  {journal} {Physical Review Letters}\ }\textbf {\bibinfo {volume} {101}},\ \href {https://doi.org/10.1103/physrevlett.101.263602} {10.1103/physrevlett.101.263602} (\bibinfo {year} {2008})\BibitemShut {NoStop}%
\bibitem [{\citenamefont {Yamaguchi}(2017)}]{Yamaguchi_GaAs_nanoresonator}%
  \BibitemOpen
  \bibfield  {author} {\bibinfo {author} {\bibfnamefont {H.}~\bibnamefont {Yamaguchi}},\ }\bibfield  {title} {\bibinfo {title} {Gaas-based micro/nanomechanical resonators},\ }\href {https://doi.org/10.1088/1361-6641/aa857a} {\bibfield  {journal} {\bibinfo  {journal} {Semiconductor Science and Technology}\ }\textbf {\bibinfo {volume} {32}},\ \bibinfo {pages} {103003} (\bibinfo {year} {2017})}\BibitemShut {NoStop}%
\bibitem [{\citenamefont {Bothner}\ \emph {et~al.}(2020)\citenamefont {Bothner}, \citenamefont {Yanai}, \citenamefont {Iniguez-Rabago}, \citenamefont {Yuan}, \citenamefont {Blanter},\ and\ \citenamefont {Steele}}]{Gary_Steele_parametric}%
  \BibitemOpen
  \bibfield  {author} {\bibinfo {author} {\bibfnamefont {D.}~\bibnamefont {Bothner}}, \bibinfo {author} {\bibfnamefont {S.}~\bibnamefont {Yanai}}, \bibinfo {author} {\bibfnamefont {A.}~\bibnamefont {Iniguez-Rabago}}, \bibinfo {author} {\bibfnamefont {M.}~\bibnamefont {Yuan}}, \bibinfo {author} {\bibfnamefont {Y.~M.}\ \bibnamefont {Blanter}},\ and\ \bibinfo {author} {\bibfnamefont {G.~A.}\ \bibnamefont {Steele}},\ }\bibfield  {title} {\bibinfo {title} {Cavity electromechanics with parametric mechanical driving},\ }\bibfield  {journal} {\bibinfo  {journal} {Nature Communications}\ }\textbf {\bibinfo {volume} {11}},\ \href {https://doi.org/10.1038/s41467-020-15389-4} {10.1038/s41467-020-15389-4} (\bibinfo {year} {2020})\BibitemShut {NoStop}%
\bibitem [{\citenamefont {Gonzalez}\ and\ \citenamefont {Lee}(2018)}]{Pi_2_angle_add}%
  \BibitemOpen
  \bibfield  {author} {\bibinfo {author} {\bibfnamefont {M.}~\bibnamefont {Gonzalez}}\ and\ \bibinfo {author} {\bibfnamefont {Y.}~\bibnamefont {Lee}},\ }\bibfield  {title} {\bibinfo {title} {A study on parametric amplification in a piezoelectric mems device},\ }\href {https://doi.org/10.3390/mi10010019} {\bibfield  {journal} {\bibinfo  {journal} {Micromachines}\ }\textbf {\bibinfo {volume} {10}},\ \bibinfo {pages} {19} (\bibinfo {year} {2018})}\BibitemShut {NoStop}%
\end{thebibliography}%


\begin{thebibliography}{2}%
\makeatletter
\providecommand \@ifxundefined [1]{%
 \@ifx{#1\undefined}
}%
\providecommand \@ifnum [1]{%
 \ifnum #1\expandafter \@firstoftwo
 \else \expandafter \@secondoftwo
 \fi
}%
\providecommand \@ifx [1]{%
 \ifx #1\expandafter \@firstoftwo
 \else \expandafter \@secondoftwo
 \fi
}%
\providecommand \natexlab [1]{#1}%
\providecommand \enquote  [1]{``#1''}%
\providecommand \bibnamefont  [1]{#1}%
\providecommand \bibfnamefont [1]{#1}%
\providecommand \citenamefont [1]{#1}%
\providecommand \href@noop [0]{\@secondoftwo}%
\providecommand \href [0]{\begingroup \@sanitize@url \@href}%
\providecommand \@href[1]{\@@startlink{#1}\@@href}%
\providecommand \@@href[1]{\endgroup#1\@@endlink}%
\providecommand \@sanitize@url [0]{\catcode `\\12\catcode `\$12\catcode `\&12\catcode `\#12\catcode `\^12\catcode `\_12\catcode `\%12\relax}%
\providecommand \@@startlink[1]{}%
\providecommand \@@endlink[0]{}%
\providecommand \url  [0]{\begingroup\@sanitize@url \@url }%
\providecommand \@url [1]{\endgroup\@href {#1}{\urlprefix }}%
\providecommand \urlprefix  [0]{URL }%
\providecommand \Eprint [0]{\href }%
\providecommand \doibase [0]{https://doi.org/}%
\providecommand \selectlanguage [0]{\@gobble}%
\providecommand \bibinfo  [0]{\@secondoftwo}%
\providecommand \bibfield  [0]{\@secondoftwo}%
\providecommand \translation [1]{[#1]}%
\providecommand \BibitemOpen [0]{}%
\providecommand \bibitemStop [0]{}%
\providecommand \bibitemNoStop [0]{.\EOS\space}%
\providecommand \EOS [0]{\spacefactor3000\relax}%
\providecommand \BibitemShut  [1]{\csname bibitem#1\endcsname}%
\let\auto@bib@innerbib\@empty
\bibitem [{\citenamefont {Agarwal}\ \emph {et~al.}(2006)\citenamefont {Agarwal}, \citenamefont {Park}, \citenamefont {Candler}, \citenamefont {Kim}, \citenamefont {Hopcroft}, \citenamefont {Chandorkar}, \citenamefont {Jha}, \citenamefont {Melamud}, \citenamefont {Kenny},\ and\ \citenamefont {Murmann}}]{Manu_electrostatic}%
  \BibitemOpen
  \bibfield  {author} {\bibinfo {author} {\bibfnamefont {M.}~\bibnamefont {Agarwal}}, \bibinfo {author} {\bibfnamefont {K.~K.}\ \bibnamefont {Park}}, \bibinfo {author} {\bibfnamefont {R.~N.}\ \bibnamefont {Candler}}, \bibinfo {author} {\bibfnamefont {B.}~\bibnamefont {Kim}}, \bibinfo {author} {\bibfnamefont {M.~A.}\ \bibnamefont {Hopcroft}}, \bibinfo {author} {\bibfnamefont {S.~A.}\ \bibnamefont {Chandorkar}}, \bibinfo {author} {\bibfnamefont {C.~M.}\ \bibnamefont {Jha}}, \bibinfo {author} {\bibfnamefont {R.}~\bibnamefont {Melamud}}, \bibinfo {author} {\bibfnamefont {T.~W.}\ \bibnamefont {Kenny}},\ and\ \bibinfo {author} {\bibfnamefont {B.}~\bibnamefont {Murmann}},\ }\bibfield  {title} {\bibinfo {title} {Nonlinear characterization of electrostatic mems resonators},\ }in\ \href {https://doi.org/10.1109/FREQ.2006.275380} {\emph {\bibinfo {booktitle} {2006 IEEE International Frequency Control Symposium and Exposition}}}\ (\bibinfo {year} {2006})\ pp.\ \bibinfo {pages} {209--212}\BibitemShut {NoStop}%
\bibitem [{\citenamefont {Kaajakari}\ \emph {et~al.}(2004)\citenamefont {Kaajakari}, \citenamefont {Mattila}, \citenamefont {Oja},\ and\ \citenamefont {Seppa}}]{Kaajakari_nonlinearity}%
  \BibitemOpen
  \bibfield  {author} {\bibinfo {author} {\bibfnamefont {V.}~\bibnamefont {Kaajakari}}, \bibinfo {author} {\bibfnamefont {T.}~\bibnamefont {Mattila}}, \bibinfo {author} {\bibfnamefont {A.}~\bibnamefont {Oja}},\ and\ \bibinfo {author} {\bibfnamefont {H.}~\bibnamefont {Seppa}},\ }\bibfield  {title} {\bibinfo {title} {Nonlinear limits for single-crystal silicon microresonators},\ }\href {https://doi.org/10.1109/jmems.2004.835771} {\bibfield  {journal} {\bibinfo  {journal} {Journal of Microelectromechanical Systems}\ }\textbf {\bibinfo {volume} {13}},\ \bibinfo {pages} {715–724} (\bibinfo {year} {2004})}\BibitemShut {NoStop}%
\end{thebibliography}%

\end{document}



\title{\textbf{Supplementary Materials: Selective Parametric Amplification of Degenerate Modes in Electrostatically Transduced Coupled Beam Resonators} 
}%

\author{Vishnu Kumar}
\author{Nishta Arora}
\author{Bhargavi B.A.}
\author{Akshay Naik}
\author{Saurabh A. Chandorkar}%
\affiliation{%
Centre for Nano Science and Engineering, Indian Institute of Science, 560012 Bengaluru, India.\\
}%

\date{\today}

\maketitle

\section{Device Fabrication}\label{Appen_I}

\begin{figure}[!h]
    \centering
    \includegraphics[width =\linewidth]{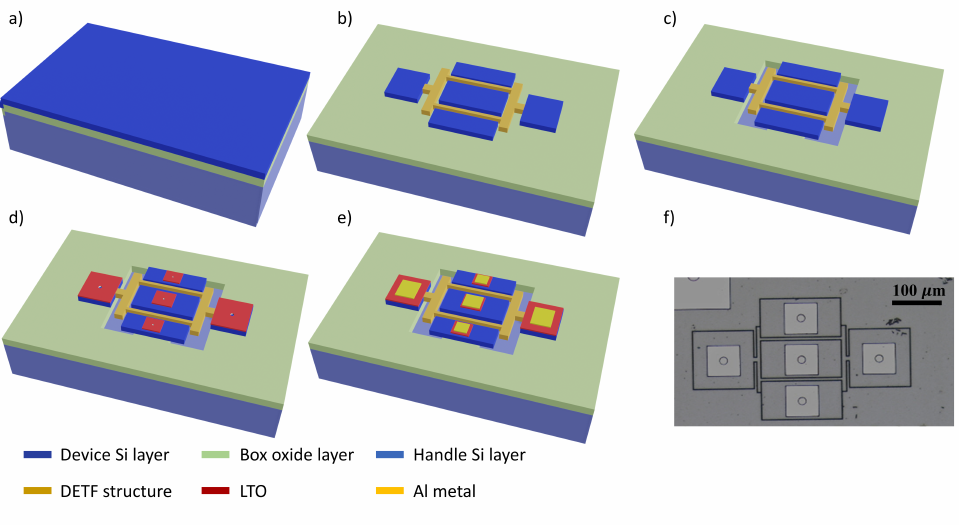}
    \caption{Device fabrication: (a) silicon-on-insulator wafer, (b) defined and etched DETF structure using lithography and DRIE, (c) Vapor HF to release the device structure from the bottom oxide layer, (d) LTO deposition in LPCVD and subsequent RIE to etch the oxide layer for device layer contact opening, (e) Al metal deposition and etch, and (f) SEM image of the DETF resonator}
    \label{fab}
\end{figure}

The device fabrication begins with a silicon-on-insulator (SOI) wafer consisting of a 20 $\mu m$ thick boron-doped device layer and a 2 $\mu m$ buried oxide (BOX) layer, as shown in \autoref{fab}a. First, lithography defines the resonator structure, preceded by the deposition of PECVD oxide as a hard mask for Deep Reactive Ion Etching (DRIE) of the device layer, as illustrated in \autoref{fab}b. After etching the device layer, the oxide mask is removed, and the device undergoes vapor HF etching to selectively remove the BOX layer, releasing the structure from the substrate (\autoref{fab}c).

Low-Temperature Oxide (LTO) is then deposited via LPCVD to non-conformally deposit the DRIE-etched trenches, serving as a protective layer to prevent metal-induced shorting. A second lithography step patterns the LTO, followed by reactive ion etching (RIE) to expose contact areas on the device layer (\autoref{fab}d). Subsequently, an aluminum metal layer is deposited and annealed to ensure good electrical contact. Finally, the aluminum is patterned and etched, and the device is released by removing the top LTO layer (\autoref{fab}e). The final fabricated device is shown in the SEM image in \autoref{fab}f.

\section{Analytical Modeling}\label{Appen_II}

The equation of motion for DETF is:

\begin{equation}\label{eqn 3a}
    m\Ddot{x_1}+c\dot{x_1}+kx_1+k_c(x_1-x_2)=f_1
\end{equation}
\begin{equation}\label{eqn 3b}
    m\Ddot{x_2}+c\dot{x_2}+kx_2+k_c(x_2-x_1)=f_2
\end{equation}
where, $m, c, k=$mass, damping, and stiffness of the resonating beam, $k_c=$ stiffness of the coupling beam, and $g=$gap between the electrode and resonating beam. 
Considering in-phase mode: $x_1+x_2=u$, 
and out-phase mode: $x_1-x_2=v$

The \autoref{eqn 3a} $\&$ \autoref{eqn 3b} become:

\begin{equation}\label{eqn 3c}
    m\Ddot{u}+c\dot{u}+ku=f_1+f_2
\end{equation}
\begin{equation}\label{eqn 3d}
    m\Ddot{v}+c\dot{v}+kv+2k_cv=f_1-f_2
\end{equation}

The natural frequencies of the two degenerate modes are:
In-phase mode: $\omega_1=\sqrt{k/m}$, and out-phase mode: $\omega_2=\sqrt{(k+2k_c)/m}$

\begin{equation}\label{eqn 4_1}
    m\Ddot{u}+c\dot{u}+ku=f_1+f_2
\end{equation}
\begin{equation}\label{eqn 4_2}
    m\Ddot{v}+c\dot{v}+kv+2k_cv=f_1-f_2
\end{equation}

The forces $f_1$ and $f_2$ are acting on the resonating beams 1 and 2 respectively. As the actuation is applied at beam 1, the force has components from the actuation electrode ($V_{ac}$) and middle electrode (being zero potential).

\begin{equation}\label{eqn 4_3}
    f_1=\dfrac{1}{2}\dfrac{\epsilon A V_{dc}^2}{(g-x_1)^2}-\dfrac{1}{2}\dfrac{\epsilon A (V_{dc}-V_{ac})^2}{(g+x_1)^2}
\end{equation}

\begin{equation}\label{eqn 4_4}
    f_2=\dfrac{1}{2}\dfrac{\epsilon A V_{dc}^2}{(g-x_2)^2}-\dfrac{1}{2}\dfrac{\epsilon A V_{dc}^2}{(g+x_2)^2}
\end{equation}

The force exerted for in-phase mode is as:

\begin{equation}\label{eqn 4_5}
    f_1+f_2=\dfrac{\epsilon A}{g^2}V_{dc}V_{ac}+\dfrac{2\epsilon A}{g^3}V_{dc}^2(x_1+x_2)+\dfrac{4\epsilon A}{g^5}V_{dc}^2(x_1^3+x_2^3)
\end{equation}

The force exerted for out-phase mode is as:

\begin{equation}\label{eqn 4_6}
    f_1-f_2=\dfrac{\epsilon A}{g^2}V_{dc}V_{ac}+\dfrac{2\epsilon A}{g^3}V_{dc}^2(x_1-x_2)+\dfrac{4\epsilon A}{g^5}V_{dc}^2(x_1^3-x_2^3)
\end{equation}

\autoref{eqn 4_5} and \autoref{eqn 4_6} show the electrostatic nonlinearity due to the electrostatic transduction. 

As, $x_1+x_2=u$ and $x_1-x_2=v$, the $x_1x_2=\dfrac{u^2-v^2}{4}$. The \autoref{eqn 4_5} and \autoref{eqn 4_6} can be simplified in u and v terms as:

\begin{equation}\label{eqn 4_7}
    f_1+f_2=f_a cos(\omega t)+f_bu+f_c(3v^2u+u^3)
\end{equation}
\begin{equation}\label{eqn 4_8}
    f_1-f_2=f_a cos(\omega t)+f_bv+f_c(3u^2v+v^3)
\end{equation}

Here, $f_a=\dfrac{\epsilon A V_{dc}V_{ac,0}}{g^2}$, $f_b=\dfrac{2\epsilon A}{g^3}V_{dc}^2$, and $f_c=\dfrac{4\epsilon A}{g^5}V_{dc}^2$

Measurement performed for the out-phase mode can have a negligible amplitude of in-phase mode since their resonance frequencies are sufficiently apart, i.e., measurement carried out at $u$ will have $v \sim 0$, so $u^2v \sim 0$ and vice versa.

Now, the equation of motion \autoref{eqn 4_1} and \autoref{eqn 4_2} for out-phase and in-phase modes can be written as: 

\begin{equation}\label{eqn 4_9}
    m\Ddot{u}+c\dot{u}+ku=f_a cos(\omega t)+f_bu+f_cu^3
\end{equation}
\begin{equation}\label{eqn 4_10a}
    m\Ddot{v}+c\dot{v}+kv+2k_cv=f_a cos(\omega t)+f_bv+f_cv^3
\end{equation}

We further recognize that the coupling spring also experiencing electrostatic nonlinearity since the coupling beam is at a potential $V_{dc}$ while inner electrode is at ground potential. Thus, the out-phase mode equation is modified as:

\begin{equation}\label{eqn 4_10b}
    m\Ddot{v}+c\dot{v}+kv+2k_{c1}v-2k_{c3}v^3=f_a cos(\omega t)+f_bv+f_cv^3
\end{equation}

Taking mechanical nonlinearity into account, the \autoref{eqn 4_9} and \autoref{eqn 4_10b} are as:

\begin{equation}\label{eqn 4_11}
    m\Ddot{u}+c\dot{u}+k_1u+k_3u^3=f_a cos(\omega t)+f_bu+f_cu^3
\end{equation}
\begin{equation}\label{eqn 4_12}
    m\Ddot{v}+c\dot{v}+k_1v+k_3v^3+2k_{c1}v-2k_{c3}v^3=f_a cos(\omega t)+f_bv+f_cv^3
\end{equation}


The electrostatic nonlinearity of the resonating beam is approximately equal to the mechanical nonlinearity of the resonating beam at the dc bias of 50 V by calculation \cite{Manu_electrostatic}. The forcing coefficient of electrostatic nonlinearity is $\sim 10^{13}$. The mechanical nonlinearity of the resonating beam $k_3$ can be obtained through the strain profile \cite{Kaajakari_nonlinearity} and it's also in the order of $10^{13}$, whereas the nonlinearity through the coupling beam shows order of $10^{18}$ (the estimation is provided in supplementary section III). Comparing to the nonlinear coefficient of coupling spring ($k_{c3}$) to the resonating spring ($k_3$), the $k_3 \ll k_{c3}$. Therefore, the effective nonlinearity is dominated by the coupling spring. Thus, the equations become:

\begin{equation}\label{eqn 4_13}
    m\Ddot{u}+c\dot{u}+(k_1-f_b)u=f_a cos(\omega t)
\end{equation}
\begin{equation}\label{eqn 4_14}
    m\Ddot{v}+c\dot{v}+(k_1+2k_{c1}-f_b)v-2k_{c3}v^3=f_a cos(\omega t)
\end{equation}

Without nonlinearity, the resonance frequencies for the two-phase modes are: $\omega_1=\sqrt{\dfrac{(k_1-f_b)}{m}}$, and $\omega_2=\sqrt{\dfrac{(k_1+2k_{c1}-f_b)}{m}}$.\\

With \autoref{eqn 4_14}, it is evident that the coupling beam nonlinearity has greater effect on the out-phase mode to excite parametrically before in-phase nonlinearity hits in. 

The equation of motion due to parametric excitation on the coupling beam beyond the critical voltage to self-oscillation regime and become nonlinear is:

\begin{equation}\label{eqn 4_15}
    m\Ddot{v}+c\dot{v}+(k_1-f_b+2k_{c1}(1+\beta sin(2\omega t)))v-2k_{c3}v^3=0
\end{equation}

Similarly, applying harmonic drive on top of the parametric drive at the same electrode and changing the angle between them to get the parametric oscillation as:

\begin{equation}\label{eqn 4_16}
    m\Ddot{v}+c\dot{v}+(k_1-f_b+2k_{c1}(1+\beta sin(2\omega t))v-2k_{c3}v^3=f_a cos(\omega t+\theta)
\end{equation}

\begin{table}[!h]
    \centering
    \begin{tabular}{|c|c|} \hline
        $\textbf{Parameters}$ & $\textbf{Value}$\\ \hline
        
         $m_{eff}$ &  $3.28E-8 \ [kg]$\\ \hline
         $A$ &  $4.9E-9 \ [m^2]$\\ \hline
         $g$ &  $2.9E-6 \ [m]$\\ \hline
         $\epsilon$ &  $1.035E-10 \ [F/m]$\\ \hline
         $V_{dc}$ &  $50 \ [V]$\\ \hline
         $\omega_1$ &  $766.1 \ [kHz]$\\ \hline
         $\omega_2$ &  $791.75 \ [kHz]$\\ \hline
         $Q$ &  $10450$\\ \hline
         $k_1$ &  $1.91E4 \ [N/m]$\\ \hline
         $k_3$ &  $1.03E13 \ [N/m^3]$\\ \hline
         $k_{c1}$ &  $6.55E2 \ [N/m]$\\ \hline
         $f_a$ &  $1.5E-6 \ [N]$\\ \hline
         $f_b$ &  $1.04E2 \ [N/m]$\\ \hline
         $f_c$ &  $2.47E13 \ [N/m^3]$\\ \hline
         $V_c (OP)$ &  $1.1 \ [V]$\\ \hline
         $V_c (IP)$ &  $20 \ [V]$\\ \hline
    \end{tabular}
    \caption{Device parameters}
    \label{Parameter table}
\end{table}

\section{Nonlinear parameters estimation}\label{Appen_III}

\begin{figure}[!h]
    \centering
    \includegraphics[width =\linewidth]{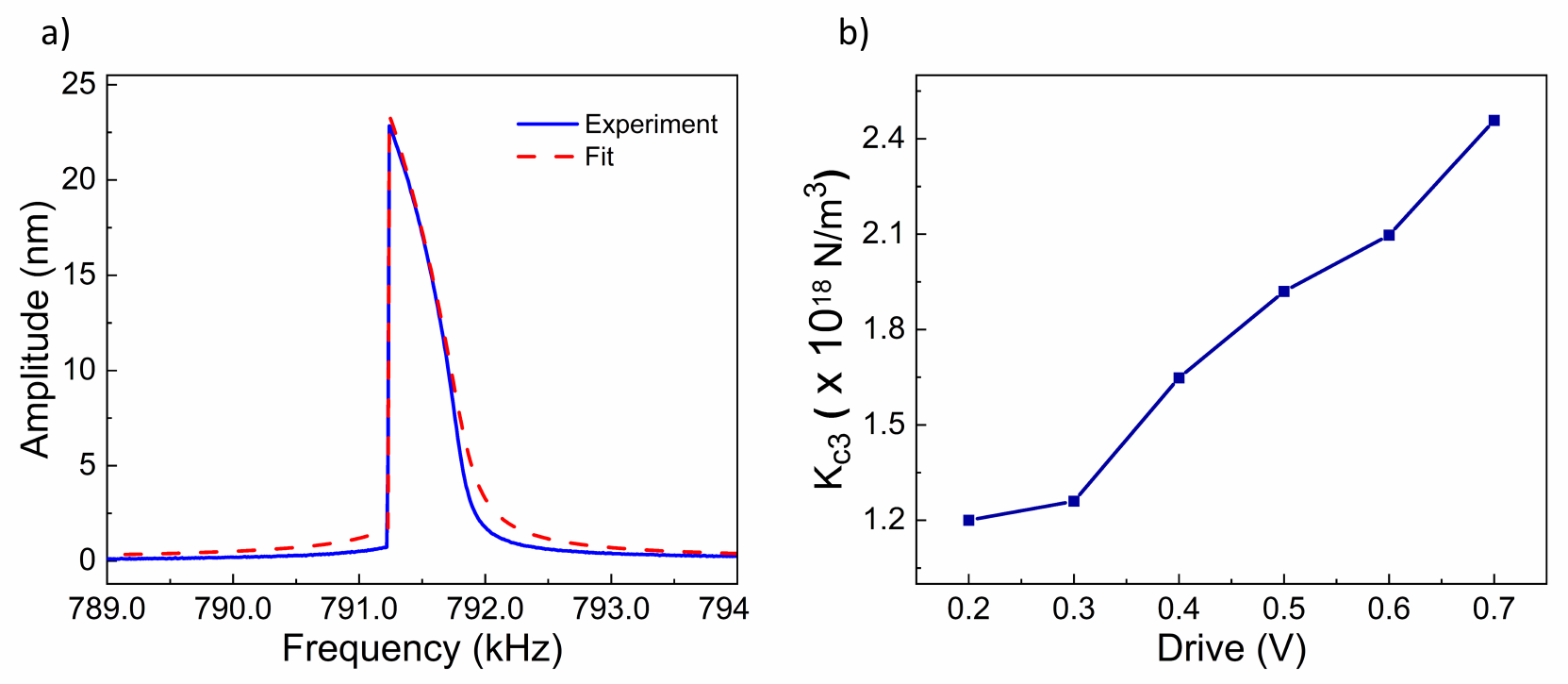}
    \caption{Nonlinear coefficient: (a) nonlinear analytical fitting of a response, the softening nonlinear coupling coefficient ($K_{c3}$) is $\sim1.26 \times 10^{18} Nm^{-3}$; (b) the variation in nonlinear coefficient with applied harmonic drive, illustrates linear behavior.}
    \label{nonlinear_fit}
\end{figure}

The parameter estimation is carried out using the \autoref{eqn 4_14} and the softening nonlinearity is accounted. The analytical modeling and response fitting is shown in the \autoref{nonlinear_fit}a and the softening nonlinear coefficient is $\sim1.26 \times 10^{18} Nm^{-3}$, whereas the higher drive shows the linear effect on the nonlinearity as shown in \autoref{nonlinear_fit}b.

The parametric coefficient ($\beta$) is estimating using the \autoref{eqn 4_15}, and the value is $1.58 \times 10^{-9} \ 1/N$.

\newpage
\bibliography{references}